\documentclass[a4paper,11pt]{article}
\pdfoutput=1

\usepackage{jinstpub}
\usepackage{xspace}

\usepackage{lineno}
\usepackage{comment}

\newcommand{\Apollo}{\texttt{Apollo}\xspace}

\newcommand{\Builder}{\texttt{EventBuilder}\xspace}
\newcommand{\DaqServer}{\texttt{DaqServer}\xspace}
\newcommand{\DataReader}{\texttt{DataReader}\xspace}
\newcommand{\DataReaders}{\texttt{DataReaders}\xspace}

\newcommand{\MirrorClient}{\texttt{MirrorClient}\xspace}
\newcommand{\MirrorServer}{\texttt{MirrorServer}\xspace}

\newcommand{\MsgLogger}{\texttt{MsgLogger}\xspace}
\newcommand{\PulserController}{\texttt{PulserController}\xspace}

\newcommand{\EleController}{\texttt{EleController}\xspace}

\newcommand{\ignore}[1]{}

\title{A data acquisition and control system for large mass bolometer arrays}

\author[a,b,1]{S.~Di~Domizio,\note{Corresponding author}}
\author[c,d]{A.~Branca,}
\author[b]{A.~Caminata,}
\author[e,2]{L.~Canonica,\note{Presently at Max Planck Institut f\"ur Physik,
F\"ohringer Ring 6, M\"unchen, D-80805, Germany}}
\author[f]{S.~Copello,}
\author[g,h]{A.~Giachero,}
\author[b,3]{E.~Guardincerri,\note{Presently at Los Alamos National Laboratory,
PO Box 1663, Los Alamos, NM 87545, USA}}
\author[a,b,j]{L.~Marini,}
\author[a,b]{M.~Pallavicini,}
\author[i]{M.~Vignati}

\affiliation[a]{Dipartimento di Fisica, Universit\`a degli Studi di Genova, Genova I-16146, Italy}
\affiliation[b]{INFN -- Sezione di Genova, Genova I-16146, Italy }
\affiliation[c]{Dipartimento di Fisica e Astronomia, Universit\`{a} di Padova, I-35131 Padova, Italy}
\affiliation[d]{INFN -- Sezione di Padova, Padova I-35131, Italy }
\affiliation[e]{INFN -- Laboratori Nazionali del Gran Sasso, Assergi (L'Aquila) I-67100, Italy}
\affiliation[f]{INFN -- Gran Sasso Science Institute, L'Aquila I-67100, Italy}
\affiliation[g]{INFN -- Sezione di Milano Bicocca, Milano I-20126, Italy}
\affiliation[h]{Dipartimento di Fisica, Universit\`{a} di Milano-Bicocca, Milano I-20126, Italy}
\affiliation[i]{INFN -- Sezione di Roma, Roma I-00185, Italy}
\affiliation[j]{Department of Physics, University of California, Berkeley, CA 94720, USA}

\emailAdd{sergio.didomizio@ge.infn.it}

\date{\today}

\abstract{

  During the last couple of decades, the use of arrays of bolometers has represented one of the leading techniques for rare events searches.
  CUORE, an array of 988 TeO$_2$ bolometers that is taking data since April 2017 at the Laboratori Nazionali del Gran Sasso (Italy), exploits the large mass, low background, good energy resolution and low energy threshold of these detectors successfully.
  Thanks to these characteristics, they could be also sensitive to low energy rare processes, such as galactic dark matter interactions.
In this paper we describe the data acquisition system that was developed for the CUORE experiment.
  Thanks to its high modularity, the data acquisition here described has been used in different setups with similar requirements, including the pilot experiment CUORE-0 and the demonstrator for the next phase of the project, CUPID-0, also taking data at LNGS.
}


\begin{document}
\maketitle
\flushbottom

\section{Introduction}\label{sec:intro}

Large mass bolometers~\cite{bolometers:enss,Pirro:2017ecr} are excellent detectors for the search of rare events, such as neutrinoless double beta decay (0$\nu\beta\beta$)~\cite{Cremonesi:2013vla} or dark matter interactions~\cite{Klasen:2015uma}.
They can be constructed with a wide variety of materials and provide low background and excellent resolution over a wide energy range, a fundamental property for the detection of rare events.
Examples of large mass bolometers using different materials can be found in the following references \cite{Armengaud:2017hit, Casali:2013zzr, Beeman:2012wz, Cardani:2012xq, Angloher:2017sft, Angloher:2016hbv, Alenkov:2015dic}.
This detection technique also allows to build large arrays of bolometers, a crucial step in order to increase the active mass of the experiments, hence enhancing their sensitivity to rare processes.

At present the experiment which brought the bolometer technique to its greatest expression in terms of size and modularity, is CUORE~\cite{Arnaboldi2004, Artusa:2014lgv} (Cryogenic Underground Observatory for Rare Events).
CUORE is an array of 988 tellurium dioxide bolometers with a total active mass of 741$\,$kg.
The experiment started taking data in April 2017 at the Laboratori Nazionali del Gran Sasso (LNGS), Italy.
Its main scientific goal is the search for 0$\nu\beta\beta$ of $^{130}$Te whose signature is a monochromatic peak centered at the Q-value of the decay (2527.5\,keV for $^{130}$Te~\cite{Redshaw:2009zz, Scielzo:2009nh, Rahaman:2011zz}).
The detector is made of 19 identical modules named ``towers'', each one consisting in 13 floors composed by four 5x5x5\,cm$^3$ TeO$_2$ bolometers.
The CUORE collaboration already demonstrated the effectiveness of the precautions taken to reduce the radioactive contaminations, measuring a background in the region of interest as low as $(0.014 \pm 0.002)$\,counts/(keV$\cdot$kg$\cdot$y), and a lower limit for the half-life of the process investigated of $T^{0\nu}_{1/2}> 1.3\times10^{25}$~yr ~\cite{Alduino:2017ehq}.
In addition to 0$\nu\beta\beta$, CUORE could also be sensitive to galactic dark matter interactions~\cite{Alessandria:2012ha,DMCUORE-0}, solar axions~\cite{Alessandria:2012mt} and supernova explosions~\cite{Amaya:2011sn}.

Despite the great success of CUORE, the sensitivity of the experiment is limited by an irreducible background due to natural radioactivity consisting in degraded surface alpha events whose energy is reconstructed in the region of interest for 0$\nu\beta\beta$.
Bolometers based on TeO$_2$ provide an excellent energy resolution, and the procedures for the crystal production and the construction of the detectors are well established and reproducible.
However, to improve the half-life sensitivity, it is necessary to increase the source mass and to exploit active particle discrimination techniques to reduce the background.
Several R\&D activities are being pursued in parallel.
Some are in the direction of increasing the source mass by means of isotope enrichment, either in tellurium~\cite{Artusa:2016mat} or other isotopes~\cite{Artusa:2016maw,Dafinei:2017xpc}.
Others investigate the possibility to perform particle discrimination using scintillating crystals, see e.g.~\cite{Pirro:2005ar,Beeman:2012jd,Beeman:2013vda,Armengaud:2017hit}.
Finally, other R\&D concern the investigation of sensitive cryogenic light detectors~\cite{Berge:2017nys,Biassoni:2015eij,Willers:2014eoa,Schaffner:2014caa,Bellini:2016lgg}, that would open the possibility to perform particle discrimination even in the non-scintillating TeO$_2$ bolometers, exploiting the Cerenkov radiation that is emitted by the 0$\nu\beta\beta$ signal and not by the $\alpha$ background.

The first large-mass experiment based on the technology of scintillating bolometers is CUPID-0 (CUORE Upgrade with Particle IDentification)~\cite{Azzolini:2018tum}. 
CUPID-0 searches for 0$\nu\beta\beta$ of $ ^{82}$Se using highly enriched scintillating crystals of ZnSe while proving the technology of scintillating bolometers to tag and reject background from alpha particles. 
It operates 26 bolometers of which 24 enriched with $^{82}$Se and 2 made out of natural Se, arranged in a tower-like structure for a total of 5 towers.
Each crystal is monitored by two germanium light detectors and is surrounded by a light reflector.
Like CUORE, also CUPID-0 is located at LNGS and has been taking data since June 2017, demonstrating already an alpha misidentification probability lower than 10$^{-6}$ above 4500\,keV and the best limit up to date on the half-life of 0$\nu\beta\beta$ decay in $^{82}$Se~\cite{Azzolini:2018dyb}.

Given the increasing number of channels to be acquired, the readout chain becomes an important aspect of the construction of bolometer arrays.
The bolometer signal is read out by the front end electronics, followed by the data acquisition (DAQ) system which has the important role of digitizing the signal, running the trigger algorithms and permanently store the acquired data to disk for an off-line analysis.
The DAQ system also controls the electronic devices and power supplies involved in the analog part of the readout chain.
The DAQ system here described, including both hardware and software components, is called \Apollo. 

After a brief recall of the bolometers operation for both heat and light detectors, we discuss the hardware and software requirements of the DAQ for a bolometric array.
\Apollo was initially developed for CUORE.
Nonetheless, the high modularity and flexibility of this data acquisition system make it possible to use it also in other experiments, regardless of the specific characteristics of the setup such as the number of channels and the bolometer characteristics (material, mass, etc).
Accordingly, we will keep the description of hardware, software and electronics control as general as possible without any explicit reference to a particular setup. 
Only when required for the comprehension of the text, we will use the DAQ of the CUORE experiment as a practical example.
Lastly we present the implementations of \Apollo and show the achieved performances in the systems that are currently being used, dedicating particular attention to the CUORE and CUPID-0 experiments.

\section{Bolometers operation}\label{sec:bolop}

The bolometric technique consists in evaluating the energy E released by particle interactions, measuring the temperature variation $\Delta$T=E/C produced by the interaction in the absorber (crystal), where C$\propto$T$^3$ is the temperature-dependent heat capacity of the crystal~\cite{bolometers:booth}. 
Typical bolometers temperatures of operation are around 10\,mK. 
After the interaction, the bolometer thermalizes with a heat sink through a weak thermal link usually made of PTFE holders.
Given an event of fixed energy, the amplitude of the signal, the rise time and the decay time depend on the heat capacity of the crystal, on the temperature of the thermal bath, on the thermal coupling between the bolometer and the thermal bath and on the connecting link to the front-end system.
The readout of the bolometer temperature is carried out using a neutron transmutation doped (NTD) germanium thermistor~\cite{PhysRevB.41.3761} glued on the crystal, whose resistance has a steep dependence on the temperature, R(T)=R$_0$exp$\mathrm{\left({T_0}/{T}\right)^{1/2}}$, where R$_0$ depends on the doping level and on the size of the sensor and T$_0$ only on the doping level. Typical values of R$_0$ and T$_0$ are of the order of a few~$\Omega$ and a few~K respectively.
The NTDs are made from germanium wafers exposed to a thermal neutron beam and then cut into small pieces whose dimensions, typically of the order of a few mm$^3$, are chosen depending on the type of bolometer on which the NTD is glued on. 
At the operating temperature of 10\,mK, the static resistance of the NTD varies from a few to hundreds of M$\Omega$, depending on the size of the NTD and on the operating conditions. 
If the thermistor is biased with a constant current, an energy release appears at its ends as a voltage drop from a few to hundreds of $\mu$V/MeV, depending on the specific characteristics of the bolometer and on the experimental conditions.

\begin{figure}[bt]
\begin{center}
\includegraphics[width=0.497\textwidth]{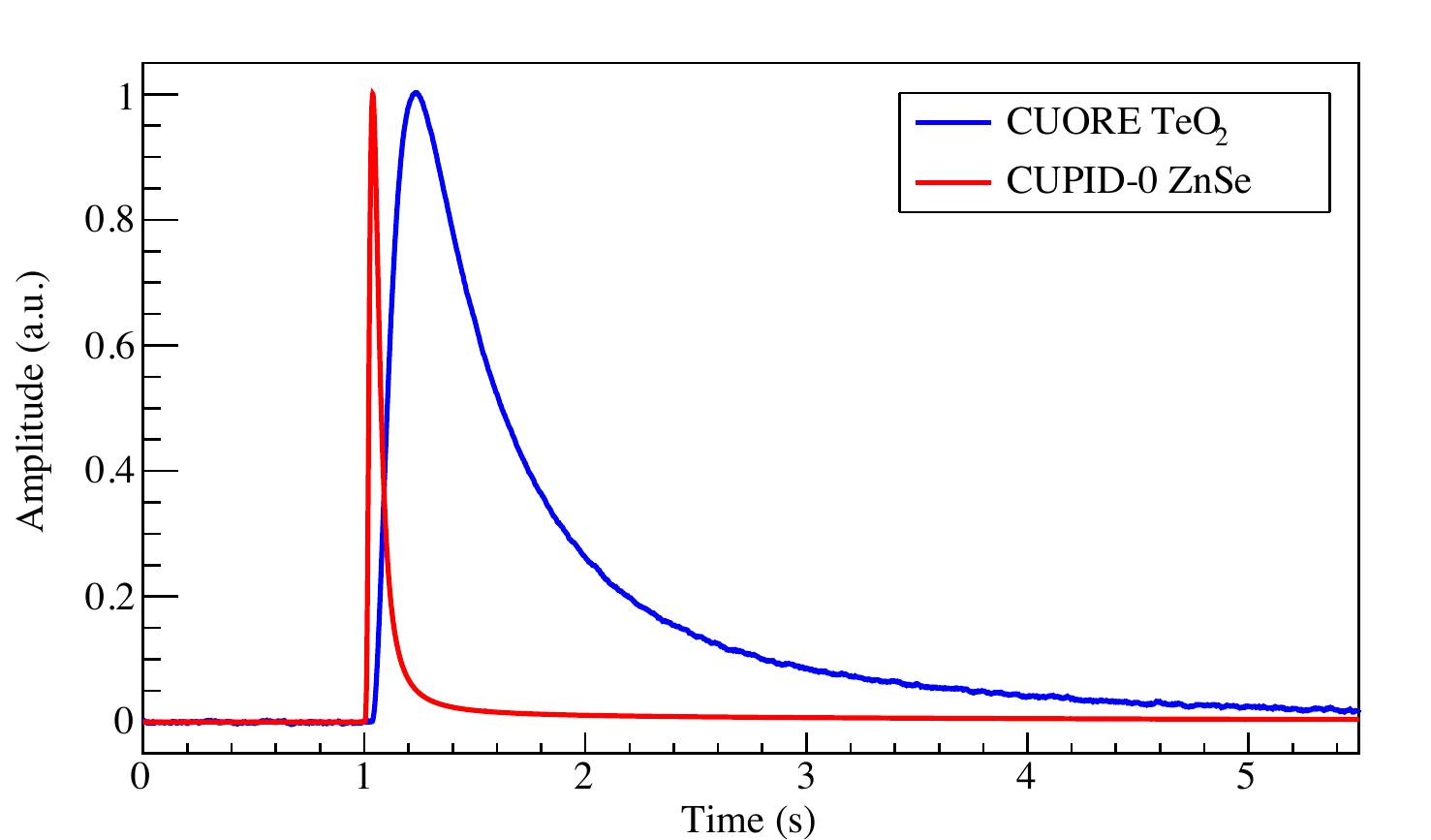}
\includegraphics[width=0.497\textwidth]{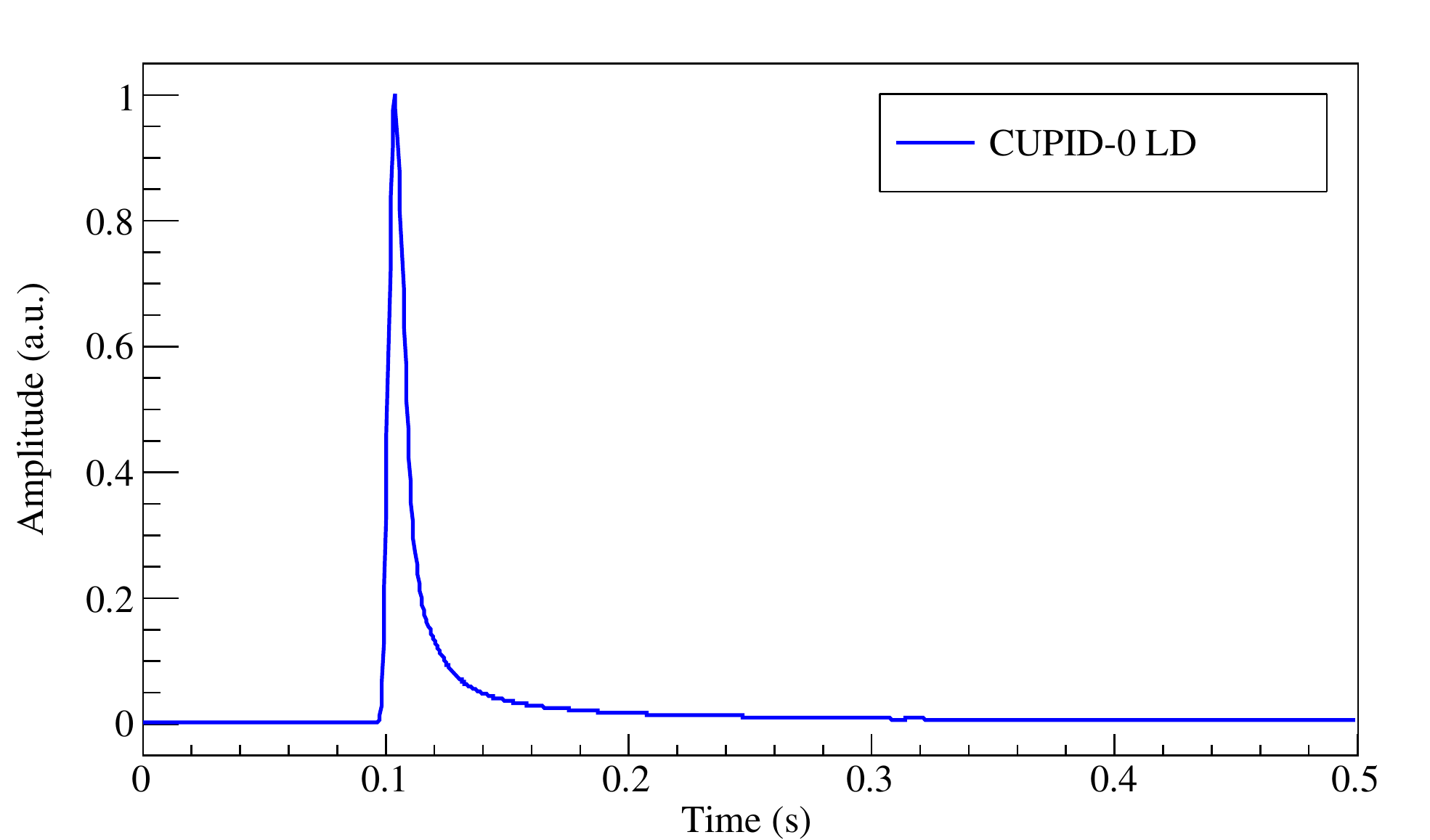}
\end{center}
\caption{Time evolution of a typical signal pulse from cryogenic detectors: on the left are reported typical pulses from CUORE TeO$_2$ crystals and from the CUPID-0 ZnSe scintillating bolometers.
  Note that the CUORE pulse has been cut to fit in the CUPID-0 event window.
  On the right a typical signal pulse from the CUPID-0 Ge LD.}
\label{fig:fig1}
\end{figure}

\Apollo is currently used with different kinds of bolometers, thus it has to be possible to tune its parameters on a channel by channel basis to ensure the best performances in all cases.
An excellent example of the versatility of \Apollo is the CUPID-0 experiment, where two different kinds of detectors are simultaneously operated and acquired.
A few-percentage of the energy released in the large-mass ZnSe absorber crystals are converted in scintillation light and measured by thin disk-shaped germanium light detectors (LD)~\cite{Beeman:2013zva}.
Given the difference in mass, the size of the NTD must take into account the difference in heat capacities: in order not to increase the heat capacity of the Ge bolometers, their NTDs must be smaller than the ones used for the main ZnSe absorber crystals.
Figure \ref{fig:fig1} reports three signal pulses that highlight the difference in the signal shape for the bolometers that we just described. 
The most evident feature is the difference in the time evolution that makes an adaptable sampling frequency and event window length necessary. 
Table \ref{tab:tab1} reports some of the main characteristics of these three kinds of bolometers and how, given the shape of their signals, \Apollo can adapt its parameters to them.

\begin{table}[h]
\centering
\begin{tabular}{ l | c  c  c }
  \hline \hline
               & CUORE  & CUPID-0 & CUPID-0 \\
               & TeO$_2$  & ZnSe       & LD \\
  \hline 	
 Crystal mass (g) & 750 & $\sim$450 &1 \\
 Responsivity ($\mu$V/MeV)&$\sim$100& 60   &10$\times$10$^3$\\

 NTD resistance (M$\Omega$) & $\sim$10$^2$ & $\sim$10 & $\sim$1 \\
 Rise time (10\%--90\%) (ms)&100 & 10  & 4  \\
 Decay time (90\%--30\%) (ms)& 400 & 40 & 8 \\

 \hline
 Sampling frequency (kHz)& 1 & 1 & 2\\
 Event window (s)& 10 & 5 & 0.5\\
 Pre-trigger (s)& 3 & 1 & 0.1 \\

  \hline \hline
\end{tabular}
\caption{Bolometer and NTD characteristics for the CUORE \cite{Alduino:2016vjd} and CUPID-0  \cite{Azzolini:2018tum} experiments. The pre-trigger window contains samples recorded before the trigger fires.
  Note that the values from the table are approximate or averaged over the available detectors.}
\label{tab:tab1}
\end{table}

Independently from the type of bolometers, the readout chain has the same structure (see Fig.~\ref{fig:fig2}): the signal readout is composed by three steps, a low noise, programmable gain amplifier, a 6-poles anti-aliasing Bessel filter and a digitizer.
In order to suppress the common mode noise, which is particularly relevant in the very first part of the readout chain, all the readout is performed in a differential configuration.
More information on the readout chain can be found in~\cite{Arnaboldi:2017aek, Arnaboldi:2010zz, Carniti:2017zkr, Arnaboldi:2015wvc, Carniti:2016rsi}.

Lastly, each detector is also provided with a very stable $\sim$100$\,$k$\Omega$ silicon resistor glued on the crystal, that is used to inject periodically a fixed amount of energy.
These heaters are driven by an ultra stable pulse generator~\cite{Arnaboldi:2003yp, Carniti:2017zkr}, producing heat signals similar to those produced by particle interactions.
The heater signals are used to compensate for small temperature instabilities by monitoring the thermal gain of the detectors and correcting it off-line to improve their energy resolution.
 In CUORE, a single pulser channel is connected in parallel to the Joule heaters of a whole detector column.
  Consequently, if the pulser fired erratically, a signal would appear in all the channels belonging to the same column and it could be easily identified.

\section{Data acquisition hardware}\label{sec:daqhardware}

The \Apollo data acquisition system is based on commercial digitizers.
This choice was done after investigating the possibility of developing custom digitizer boards~\cite{guardincerri}.
The commercial solution was adopted because it guaranteed adequate performances and at the same time allowed to reduce the costs without requiring efforts for the production.

While in some small test systems \Apollo proved to work nicely also with other boards (e.g. NI-PXI-6133), in CUORE and most other applications
the chosen digitizer board is NI-PXI-6284 from National Instruments (NI)~\cite{niboard}.
It can acquire up to 16 differential or 32 single-ended signals.
The board is provided with an 18-bit resolution analog to digital converter (ADC) and has a programmable sampling frequency up to 500\,kSamples/s.
It has a calibration circuitry to correct gain and offset errors: an internal reference signal ensures high accuracy and stability over time and temperature changes.
A multiplexer is used to digitize the signals from many channels using a single ADC.
An analog circuitry is present in the input stage of the board.
It allows to choose between differential and single-ended signals, select input range from $\pm$0.1\,V up to $\pm$10.5\,V, and toggle the status of the anti-aliasing filter (cutoff frequency at 40\,kHz or 750\,kHz).
Several analog output channels, as well as digital I/O lines are available, and  a variety of trigger signals can be imported or exported. 
The board has an internal clock with a base frequency of 80~MHz, but can also accept an external clock signal.

The sampling frequency used to digitize the data must be chosen considering the features of the signal that has to be recorded: pulses have a frequency bandwidth that extends up to about 10\,Hz.
For instance in CUORE-0 a sampling frequency of 125\,Hz was chosen because it was adequate to properly reconstruct the signals, while for CUORE 1\,kHz was selected because a more accurate pulse reconstruction could allow further studies on pulse shape of particle events. 
Pulses from light detectors are faster than those from heat sensors: they are usually digitized at 2\,kHz.

The ADC resolution is appropriate as long as it remains negligible compared to other sources of noise.
The maximum ADC input range of the NI-PXI-6284 boards is $\pm$10.5\,V, resulting in a single bit resolution of $\sim$\,80\,$\mu$V at the digitizer input, well below the RMS of the baseline of our bolometers.
Indeed the typical response of our bolometers, given by the responsivity ($\sim$100\,$\mu$V\,/\,MeV) and the front end amplification (typically 5000), is about 500\,$\mu$V\,/\,keV; therefore the ADC resolution corresponds to $\sim$\,160\,eV which is much smaller than the typical detector energy resolution (a few keV).
Similar reasoning can be made for light detectors: in this case, the response is $\sim$25\,$\mu$V\,/\,eV ($\sim$10\,nV\,/\,eV times a factor 2500 given by the front end amplification), hence the ADC resolution is less than $\sim$3\,eV, to be compared to an intrinsic baseline resolution of few tens of~eV.
Finally, the board allows to preserve the differential signal configuration that was chosen for the analog part of the readout chain, and the input voltage range matches the amplitude of the signals at the output of the Bessel filters.

\begin{figure}[bt]
\begin{center}
\includegraphics[width=1.0\textwidth]{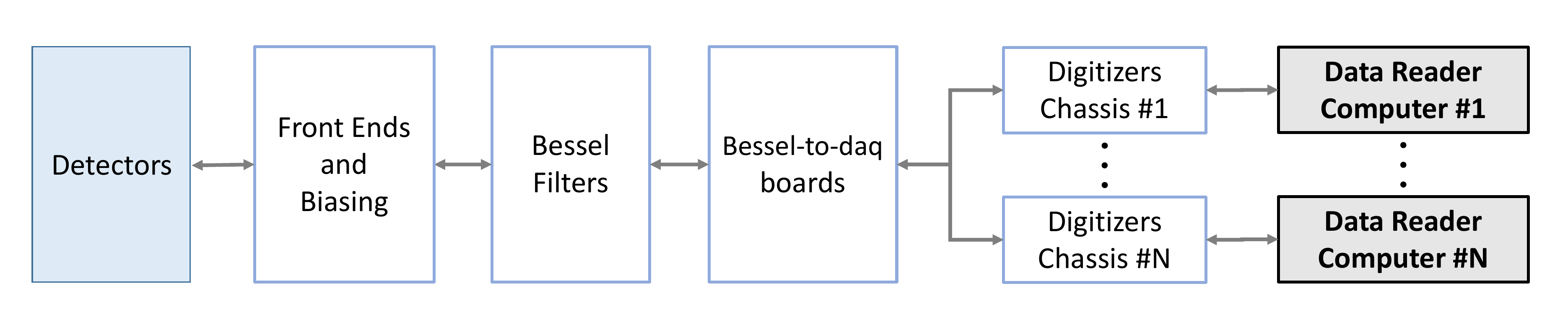}
\end{center}
\caption{Schematic representation of the readout chain of a large mass bolometer array.
  The detector signals pass through the front end electronics (which contain also the biasing circuits), the anti-aliasing Bessel filters and are finally digitized by the \Apollo digitizer boards.
  To match the different connectors and channel modularities of the Bessel and digitizer boards, a passive interface board is placed between them.}
\label{fig:fig2}
\end{figure}

The digitizer boards are enclosed in the NI PXI chassis in which the first slot hosts the controller module (PCI-PXI-8336) that is connected to a dedicated data reader computer via an optical link.
This guarantees complete isolation of the readout system from the computers power supplies.
The high modularity of this system allows integrating several chassis in the same data acquisition, as sketched in Fig.~\ref{fig:fig2}.
In particular, this kind of scalability is required in CUORE, where 64 digitizer boards (corresponding to 1024 channels) are hosted in six chassis.
The PXI chassis are provided with connectors for importing and exporting a clock source.
Then the clock signal can be delivered through the PXI bus to all the digitizer boards contained in the chassis.
This feature is exploited to connect the chassis in daisy chain so that all the digitizer boards share the same clock source.
The delay of the clock between the different chassis, introduced by the length of the cable used to connect them, can be easily evaluated and corrected for; however it is so small compared to the typical response time of the bolometers and to the digitizers sampling period, that it can be safely neglected.

Since the connectors of the digitizer boards are incompatible with the 12~channel D-Sub connectors coming from the Bessel filters, a series of custom interface boards have been designed to remap two 12-channel connectors into three 8-channel connectors compatible with the NI digitizers.
These Bessel-to-DAQ interface boards also allow accessing the I/O digital lines of the digitizer boards through 26 pole IDC connectors.
Eventually the digitized data produced by every chassis are transmitted by the optical link to the computers where they are processed and stored. 
At the end of the run, which typically lasts 24 hours, data are automatically copied to two independent server farms.
  After the file integrity has been verified with md5 checksum, data are automatically deleted from the DAQ computers.
  To guarantee the availability of the system even in case of connection problems to the remote servers, the DAQ computers can store up to about 20 days of data before saturating the local storage.

\section{Data acquisition software}\label{daqsoftware}
\begin{figure}[bt]
\begin{center}
\includegraphics[width=1.0\textwidth]{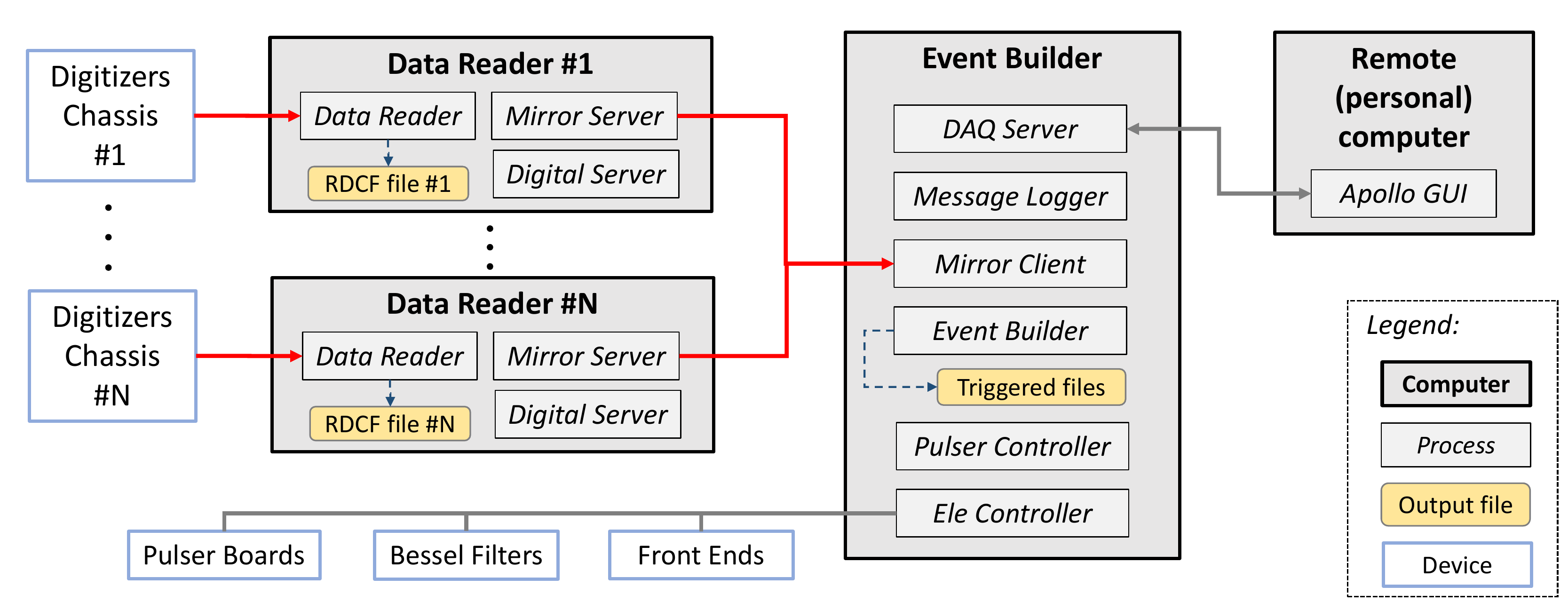}
\end{center}
\caption{Schematic representation of the data acquisition software, including host computers (Sec.~\ref{sec:daqhardware}), processes (Secs. ~\ref{daqsoftware} and~\ref{sec:elecontrol}), and output files.}
\label{fig:fig3}
\end{figure}

The \Apollo software is composed of several independent processes, possibly running on multiple computers and communicating with each other over TCP/IP to ensure the scalability and modularity of the system.
An overview of the \Apollo processes is sketched in Fig.~\ref{fig:fig3}.
The \Apollo software is written in the C++ programming language and is compiled against the NI and ROOT~\cite{Brun:1997pa} libraries.
The core of the system consists of several processes (one for each chassis controller computer) that read the data from the digitizer boards (\DataReader), and a process that collects the continuous data streams from all the data readers and builds the events that are used for off-line analysis (\Builder).
Once they are read from the digitizers, the data are made accessible to the \Apollo processes by means of shared memories that contain circular buffers. 

The \DataReader is in charge of transferring the data from the digitizers into the shared memories, running the trigger algorithms and storing the acquired waveforms into ROOT files.
As a consequence of the low frequency of the bolometer signal, the sampling frequency of the digitizers can be so low that the triggers can be implemented in the software.
This is simpler and more flexible than a hardware implementation.
For the same reason, saving the continuous stream of the bolometers waveforms to files for off-line processing can be comfortably afforded. 
Storing the continuous data makes it possible to re-trigger them if a more complex trigger algorithm is developed or for further studies on data not allowed by the triggered events.
This feature has already been successfully applied to data acquired in the past, as described in more detail in Sec.~\ref{sec:applications}.
The \DataReader can run up to four trigger algorithms in parallel on each channel.
When a trigger is found, a flag is set in the data stream that is then retrieved by the \Builder.
Which trigger algorithms to enable, as well as their specific configuration parameters, can be selected independently for each channel, to meet the intrinsic inhomogeneity in the response of bolometers.
There are usually at least two trigger algorithms running in parallel: a \textit{noise} trigger and a \textit{signal} trigger. 
The former is a simple random trigger, which fires with a given probability per unit of time, and is designed to acquire part of the waveform that, apart from accidental coincidences, is devoid of pulses.
Events associated to noise triggers are used to monitor the detector stability and to evaluate, for each run and channel, the average noise power spectrum used at a later stage of the analysis~\cite{Gatti:1986cw,Radeka:1967}.
The \textit{signal} trigger fires whenever a pulse is detected in the detectors' waveforms.
More information on the signal trigger algorithms is in Sec.~\ref{sec:trigger}.
We observe that running the triggering and event-building routines online is an optional feature of \Apollo, and that in principle the same tasks could be performed offline, starting from the continuous data.
  Either choice could be made based on the particular needs of the experimental setup where \Apollo is being used.
  For example, in CUORE the triggering and the event building algorithms are run online  because they allow to monitor the detector stability with short latency.

The \Builder searches the shared memories for trigger flags set by the \DataReader and builds the corresponding events.
This process is an instance of the off-line event reconstruction and analysis software, which reads the trigger flags from the shared memories instead of reading them from files.
The advantage of sharing the same software for online and off-line event reconstruction is twofold.
It guarantees homogeneity in data handling and format between DAQ and analysis routines, and it makes it possible to have some pre-analysis routines run directly in the \Builder.
The latter has the purpose of calculating quantities such as trigger rate, amplitude spectrum, baseline RMS that can be used to monitor the detector performance with low latency.

Events produced by the \Builder are composite objects based on the ROOT data analysis framework.
They contain information about the time, the bolometer channel that caused the trigger and the kind of trigger (signal, noise or pulser).
In case of signal or noise triggers, the events contain information about the specific trigger that fired, while for pulser triggers they contain information about the configuration parameters of the fired pulse.
Besides the primary trigger that caused the construction of the event, a list of possible secondary triggers detected in a configurable time window around the primary trigger is also present.
The events also contain the bolometer waveforms in a configurable time window around the main triggers.
The waveform associated with the primary trigger can be accompanied by other waveforms from other bolometer channels that are geometrically or logically close to the one that caused the trigger.
However, to avoid data duplication, the waveforms are removed from the events before they are saved to files.
Then in the analysis software, the bolometer waveforms are retrieved at run-time from the continuous data files when needed.
All the information that is not event-based is saved in a PostgreSQL database. The database is designed to be a unique container for information about the apparatus mapping, bolometers characteristics and position.
The database is used by \Apollo to determine the configuration to be used in a given measurement (mapping, digitizers, trigger, event building and pulser configuration)  and to keep track of the acquired measurements (start and stop time, measurement type, etc.).
The same database is also used by the off-line analysis software to retrieve the information about the acquired measurements and to store other non-event based quantities that are evaluated in the various steps of the data analysis flow.

The final part of this section gives a brief overview of the other processes that are involved in the data acquisition with \Apollo.
The \DataReader-\Builder mechanism is controlled by the \DaqServer process, which has control over the whole run including starting and stopping the measurements and monitor the status of the system during acquisition.
The user can communicate with the \DaqServer using either a dedicated graphical interface or the command line.
The graphical interface includes a real-time display of the data and its noise power spectrum for a maximum of 4 channels at the time using a software oscilloscope. 
In case the \DataReader and the \Builder do not run on the same computer, a set of dedicated processes
mirror the shared memory content from the \DataReader to the \Builder.
This process is implemented using a server-client interface.
A \MirrorServer runs in each data reader computer and the \MirrorClient runs in the event builder computer.
The \EleController process, described in detail in Sec.~\ref{sec:elecontrol},  communicates with the electronic devices of the apparatus to set their configuration and monitor their status.
The \PulserController process periodically fires thermal calibration pulses in the bolometers.
These pulses are generated by the electronic boards described in Sec.~\ref{sec:bolop}.
The firing of pulses happens in two steps.
In the first step, the \EleController is exploited to configure the pulser boards by means of a CAN bus link, preparing them to fire.
In the second step, a trigger signal is sent synchronously to the pulser boards through an optical signal and to the digital lines of the NI digitizers. 
In this way, pulser events can be easily identified and flagged in the acquired data.
All the processes interact via TCP/IP with the \MsgLogger: a server process that is in charge of collecting log messages from the active processes running in all the \Apollo computers, and keep track of their execution status.
Besides the processes described above, a set of scripts and routines exist to perform operations of several kinds.
  We give below a short and not exhaustive description of the most relevant ones.
  A first category of scripts are for performing tasks that involve a large number of detector channels, such as the measurement of load curves and working points \cite{Alduino:2016vjd}, or updating the information contained in the database about mapping and DAQ parameters, using a spreadsheet file as input.
  Another category of scripts are for the administration of \Apollo.
  They perform the one-time configuration of the operating system that will host \Apollo, start the \Apollo servers at system boot and stop them on shutdown.
  Other scripts monitor the storage area of the \Apollo computers and make sure there is always enough free space by removing old data after verifying that they have been copied to at least two remote locations.
  Finally, a script checks that all the \Apollo processes are correctly running, while the computer resources (CPU, memory load and disk usage) are monitored using the Nagios software \cite{Nagios}, and an alert is issued in case of unexpected behavior.

\section{Trigger algorithms}\label{sec:trigger}

\Apollo contains a software infrastructure that easily allows to plug in new trigger algorithms as they are implemented.
Besides the random trigger algorithm used to acquire noise samples, at present two trigger algorithms exist with the purpose of detecting particle signals.
The first algorithm, called ``\textit{derivative trigger}'' or DT, fires when the signal derivative stays above threshold for a certain amount of time.
The algorithm works on the rising front of the signal and its behavior is determined by four parameters: a derivative threshold, a time interval over which the derivative is to be evaluated, a time-over-threshold and a post-trigger dead time.
With this algorithm reasonably low thresholds can be achieved (see Sec.~\ref{sec:applications}); moreover its implementation and configuration are straightforward, and for this reason it was used in most applications of \Apollo.

The other signal trigger is called ``\textit{optimum trigger}'' or OT for short.
The algorithm, described in detail in~\cite{DiDomizio:2010ph}, consists in running a simple threshold trigger on a waveform that has been previously processed with the ``\textit{optimum filter}'' technique~\cite{Radeka:1967,Gatti:1986cw}.
The OT can achieve thresholds comparable to the RMS of the detectors' baseline, which can be as low as a few keV in large mass bolometers.
However its implementation is demanding with respect to computational resources, and its configuration is more complicated than for DT.
For these reasons the OT was used in online measurements only recently in CUORE, and its past use was limited to off-line re-triggering of previously acquired data (see Sec.~\ref{sec:applications}).
Compared to the algorithm described in~\cite{DiDomizio:2010ph}, running OT online requires further precautions to make sure that the required computing resources are enough to keep up with the rate at which the data are being acquired.
The pursued approach consists in running the OT algorithm on down-sampled data.
To avoid aliasing, the down-sampling is performed after a digital recursive time-domain Chebyshev filter is applied, a good compromise between processing speed and filtering quality.
This approach is sustainable as long as the bandwidth of the down-sampled and Chebyschev-filtered data is large enough to contain the signal bandwidth.
The down-sampling factor and the Chebyshev filter parameters are configurable so that they can be adapted to the needs of the specific setup that is being considered.
For example in CUORE, where the signal bandwidth extends up to $\sim$10\,Hz, the waveforms are digitized at 1\,kHz while the OT algorithm is run on data sampled to 125\,Hz, corresponding to a down-sampling factor of 8.
The configuration of OT requires the knowledge of the ideal detector response and noise power spectrum of each channel.
Once these quantities are known, the thresholds can be defined as a multiple of the filtered noise resolution, $\sigma_f$.
Typically a threshold value of 3\,$\sigma_f$ is chosen for all channels to guarantee a noise rejection power of 99.86\%, corresponding to a signal fake rate of $f_{n} = 0.5 \cdot f_{s} \cdot (1-erf(3/\sqrt{2}))/D_{F} \simeq 1.4\cdot10^{-3}\,f_{s}/D_{F}$, where $f_{s}$ is the sampling frequency and $D_{F}$ the decimation factor.

\section{Electronics control}\label{sec:elecontrol}

\Apollo includes a process, called \EleController, that serves as a single access point for the control and configuration of the electronic devices involved at various levels in the process of the data acquisition.
The \EleController represents an abstraction layer between the low-level communication protocols and interfaces and the high-level control commands used by the other data acquisition components and user applications.
The \EleController is built on a software framework that allows to easily implement communication with any device and interface.
At present the communication with the output TTL digital lines of the \Apollo NI digitizer boards, the front end electronics~\cite{Arnaboldi:2017aek}, anti-aliasing Bessel filters~\cite{Arnaboldi:2010zz}, pulser boards~\cite{Carniti:2017zkr} and power supplies~\cite{Arnaboldi:2015wvc,Carniti:2016rsi} used in CUORE and CUPID-0 are implemented.
The handled communication interfaces include a CAN bus (used by most devices), RS232 and GPIB.

\begin{figure}[bt]
\begin{center}
\includegraphics[width=0.9\textwidth]{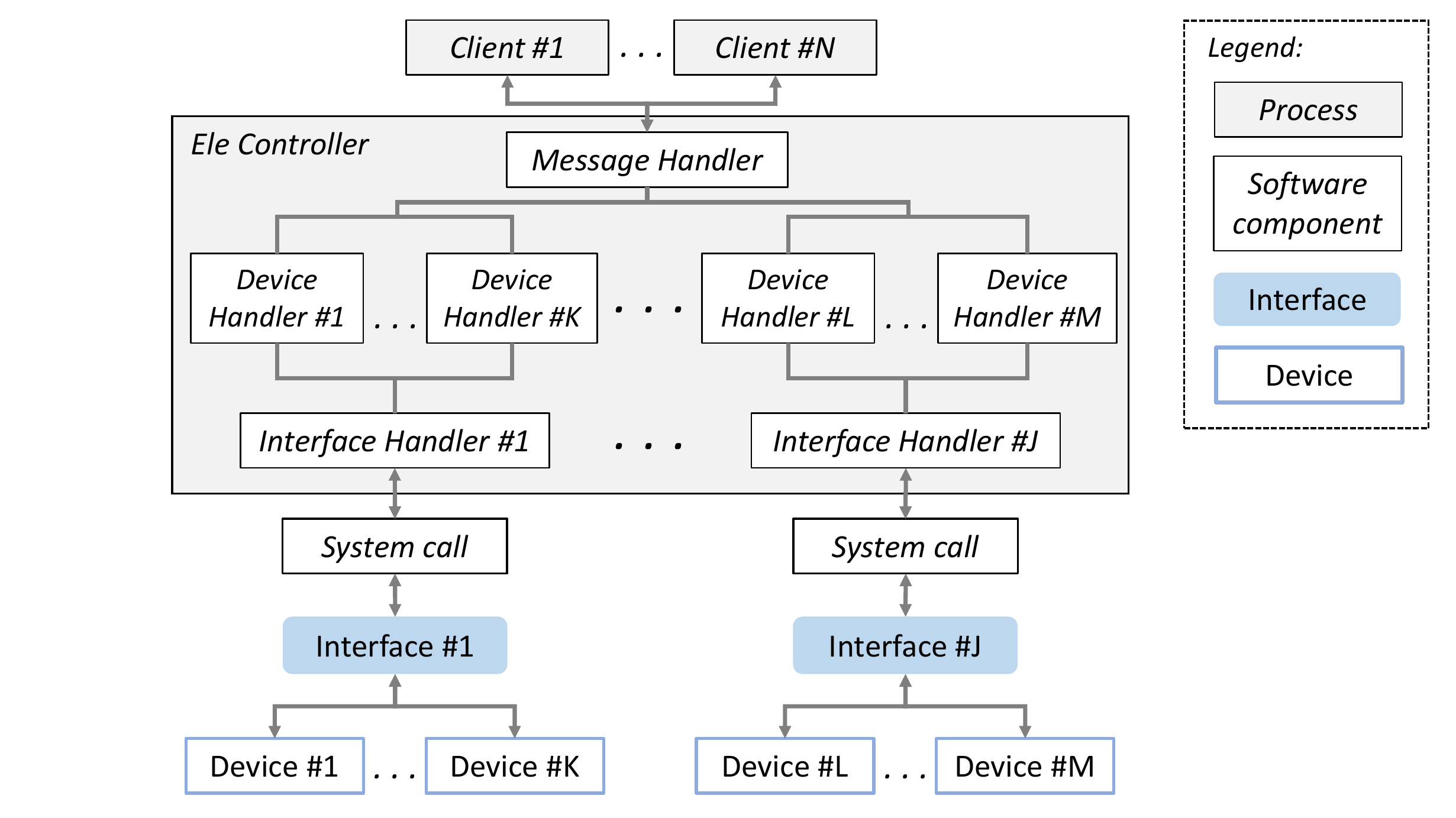}
\end{center}
\caption{Schematic representation of the \EleController process, which acts as an abstraction layer and provides concurrent communication between multiple high-level user applications and multiple low-level communication interfaces.}
\label{fig:fig4}
\end{figure}

A sketch of the \EleController infrastructure and communication mechanism is shown in Fig.~\ref{fig:fig4}.
At a high level, the data acquisition components and user applications exchange messages with the \EleController by means of a network client implementing a custom communication protocol.
The messages sent by the network client include an identifier of the target device type, an addressing part and the actual operation to be performed (typically reading or writing a parameter).
Messages can be addressed either by the physical address of the devices, e.g., chassis number, board number, and physical channel number, or by the logical channel number that is used in the data acquisition and analysis.
The conversion between the two addressing types is stored in the database.
The answer from the \EleController is a message reporting the outcome of the requested operation, or an error message in case the requested operation failed.
At low level, the system calls that interact with the communication interfaces are wrapped into interface handler classes.
The device-specific control commands are wrapped into device handler classes that make use of the interface handler classes to access the interfaces.

The \EleController is based on a multi-threaded structure that allows handling concurrent communications between multiple devices and client applications.
Multiple client connections can be handled at the same time, and for the interfaces that allow it, the interface handlers are designed to manage concurrent communications with multiple target devices.
This concurrent communication approach significantly speeds up bulk read and write operations on multiple devices, and is particularly useful for applications like CUORE, where the large number of detector channels would make a sequential communication approach unpractical.
The multi-threaded approach is made possible by the CUORE/CUPID-0 electronics, which was developed for parallel operations.
The commands handled by the \EleController are in fact macro-instructions that are executed by microcontrollers present on each electronic device (front-end, Bessel and pulser boards).

In the final part of this section, we briefly outline the high-level applications that make use of the \EleController to perform their tasks.
The most straightforward applications using the \EleController are those that are responsible for configuring and monitoring the analog components of the bolometers readout chain.
A graphical interface allows to display and control the parameters of the front end and Bessel boards, showing one channel at a time.
A command-line program was developed to perform bulk read and write operations of all the parameters of all the detector channels at once.
The configuration to be read or written by this program is stored in a spreadsheet file where each column corresponds to a different parameter, and each row corresponds to a detector channel.
Another command-line program, also using a spreadsheet for input/output, is in charge of switching on and off the power supplies of the electronic devices, and to perform their initialization after the power on.
Besides the \PulserController mentioned in section \ref{daqsoftware}, the digital lines of the NI digitizer boards are used for several other synchronization tasks.
One of them consists in sending a digital signal periodically to all the NI chassis, with the purpose of checking that their clocks are synchronized.
More complex applications consist in using the digital lines to let the data acquisition system know when specific external events occur so that the acquired data can be processed accordingly.
One example is the procedure aimed at optimizing the relative working phase of the pulse tubes of the CUORE cryostat.
In this procedure, described in detail in~\cite{DAddabbo:2017efe}, data are acquired while scanning the possible configurations of the relative phases of the pulse tubes, and the configuration that minimizes the noise on the detectors is eventually chosen for the physics data taking.
Finally, the digital lines are also exploited in the automatic procedures for the measurement of the detectors resistances and their load curves.
In these procedures the detectors baselines are acquired while changing several parameters of the readout chain, such as the thermistors' bias current and the preamplifier gain.
The \EleController plays a crucial role in these procedures because it takes care of handling both the interaction with the readout chain and with the synchronization digital lines, and thanks to its parallel communication approach allows to perform the characterization measurements in a reasonable amount of time.

\section{Applications and performance}\label{sec:applications}

The first application of \Apollo dates back to 2008 when one of its first prototypes was used for the bolometric tests of the CUORE crystals as they were being produced and shipped to LNGS~\cite{Alessandria:2011vj}.
Since then the system was used in several other cryogenic measurements~\cite{Andreotti:2009dk, Beeman:2011yc, Beeman:2011kv, Beeman:2012ci, Alessandria:2012zp, Cardani:2012xq, Beeman:2012jd, Beeman:2012wz, Beeman:2013zva, Cardani:2013dia, Beeman:2013vda, Casali:2014vvt, Alfonso:2015wka, Artusa:2016mat, Alduino:2016vtd, Pattavina:2018nhk}.
  In the following we give a brief description of the most representative applications of \Apollo.

Besides being the most important applications from a scientific perspective, the CUORE and CUPID-0 experiments represent the most advanced use of \Apollo in terms of complexity, number of handled channels and amount of features exploited.
We remark however that \Apollo is also well suited for small experimental setups where only a few channels need to be acquired.

Including the thermometers used to monitor the temperature of various points of the apparatus, the CUORE data acquisition system handles 1012 analog and several digital channels.
In all the stages of the experiment, from the construction of the detector to the realization of the data analysis software, a significant effort was necessary 
to define and then meet the requirements that could guarantee good experimental performances for the largest possible number of detectors.
Similar requirements were also of concern for the data acquisition system, that has to guarantee proper standards in terms of usability, robustness, automation and long-term stability.
The CUORE signals are digitized by 64 NI-PXI-6284 digitizer boards that are contained in six NI-PXI-1044 chassis, each chassis containing between 10 and 13 boards.
Two rack cabinets, located outside the Faraday Cage enclosing the cryostat~\cite{Bucci:2017gew} and close to the rack cabinets containing the Bessel boards, host the chassis.
Each cabinet contains three digitizer chassis and three chassis containing the Bessel-to-DAQ interface boards described in Sec.~\ref{sec:daqhardware}.
Two flat cables connect in parallel the digital lines of two boards from each of the six chassis.
One digital line is used to share a synchronous acquisition start signal between the chassis, while the other is acquired synchronously with the analog signals and is used for several purposes, for example the flagging of events generated by the pulser boards described in Sec.~\ref{daqsoftware}.
A common 10\,MHz reference clock is shared between all the chassis and ensures they remain synchronized after the start of the data acquisition.
While this configuration might change in the future, at present the signals are digitized with a sampling frequency of 1\,kHz, after they have gone through a 6-poles anti-aliasing Bessel filter with a cutoff frequency of 120\,Hz (the signal bandwidth extends up to $\sim$10\,Hz).
The software infrastructure of the CUORE data acquisition system is distributed over six data-reader computers (one for each digitizer chassis) and one computer for event building, which also hosts the other auxiliary processes and the PostgreSQL database described in Sec.~\ref{daqsoftware}.
The data-reader computers run a 32-bit version of Scientific~Linux~CERN~\cite{scientificlinux} (the NI software drivers only support 32-bit operating systems), while the computer hosting the \Builder runs a 64-bit version of the same operating system.
The computers are hosted in a rack cabinet identical to those used to contain the digitizer chassis and are connected among each other by means of a switch.

 In the CUORE implementation of \Apollo, the current trigger settings allow us to sustain a trigger rate up to a few hundreds of Hz integrated over all channels and all the trigger types, even if during standard data taking the trigger rate barely reaches 10\,Hz and less than 100\,Hz during calibration.
 With the current sampling frequency and by saving both continuous and triggered files we produce on the order of 100\,GB/day of data, mostly dominated by the continuous files which correspond to more than 99\% of the total data saved on disk.
In CUORE we demonstrated the possibility to run DT and OT in parallel, giving the opportunity to have an immediate feedback on the data with both trigger algorithms without recurring to re-triggering, discussed later on in this section.  
Finally, as already mentioned in Sec.~\ref{sec:elecontrol}, we demonstrated that we can handle the configuration of a thousand of electronic channels efficiently, exploiting tools that allow to completely automate the most frequent detector characterization and initialization procedures, such as writing and reading of electronic parameters, and reading of the NTD resistances. 
We use these tools also in other implementations of \Apollo, but in CUORE we can make the most of the processes parallelization. 

The CUPID-0 data acquisition system is the most advanced application of \Apollo in which also cryogenic light detectors are being acquired.
Including thermometers, LD and ZnSe, the total number of channels handled by the data acquisition system is 65.
Besides demonstrating the capability of the system of adapting to very different requirements, this is also relevant in view of a possible future upgrade of CUORE in which background discrimination by means of light detectors could be ported to a ton-scale experiment.
Given that the light detectors have faster signals compared to the ZnSe absorbers, CUPID-0 channels have different sampling rate and different event window associated to each trigger. 
We refer to Sec.~\ref{sec:bolop} for more details on this topic.
The signals are acquired by 7 NI-PXI-6284 digitizer boards, hosted in an NI-PXI-1044 chassis.
Consisting of a single-chassis and single-computer, the CUPID-0 data acquisition system is more straightforward than that of CUORE: no synchronization among multiple chassis is needed, and the software mechanism for replicating over the network the shared memories containing the acquired data (see Sec.~\ref{daqsoftware}) can be avoided as well.
Since the presence of the light channel is meant for particle discrimination of events in the heat channel facing it, \Apollo provides a mechanism to easily use this correspondence in the analysis phase. 
For each channel we can indeed indicate in the database a list of so called ``side channels'' that in this case coincide with the LD facing that ZnSe crystal: during the triggering phase every time that a trigger is found for the main channel, also an event in the side channel is created. 
During the analysis the side pulses, events belonging to the side channels, can be loaded and fed to the analysis chain, independently from the efficiency of the trigger or the energy evaluation.

\begin{figure}[bt]
\begin{center}
\includegraphics[width=0.8\textwidth]{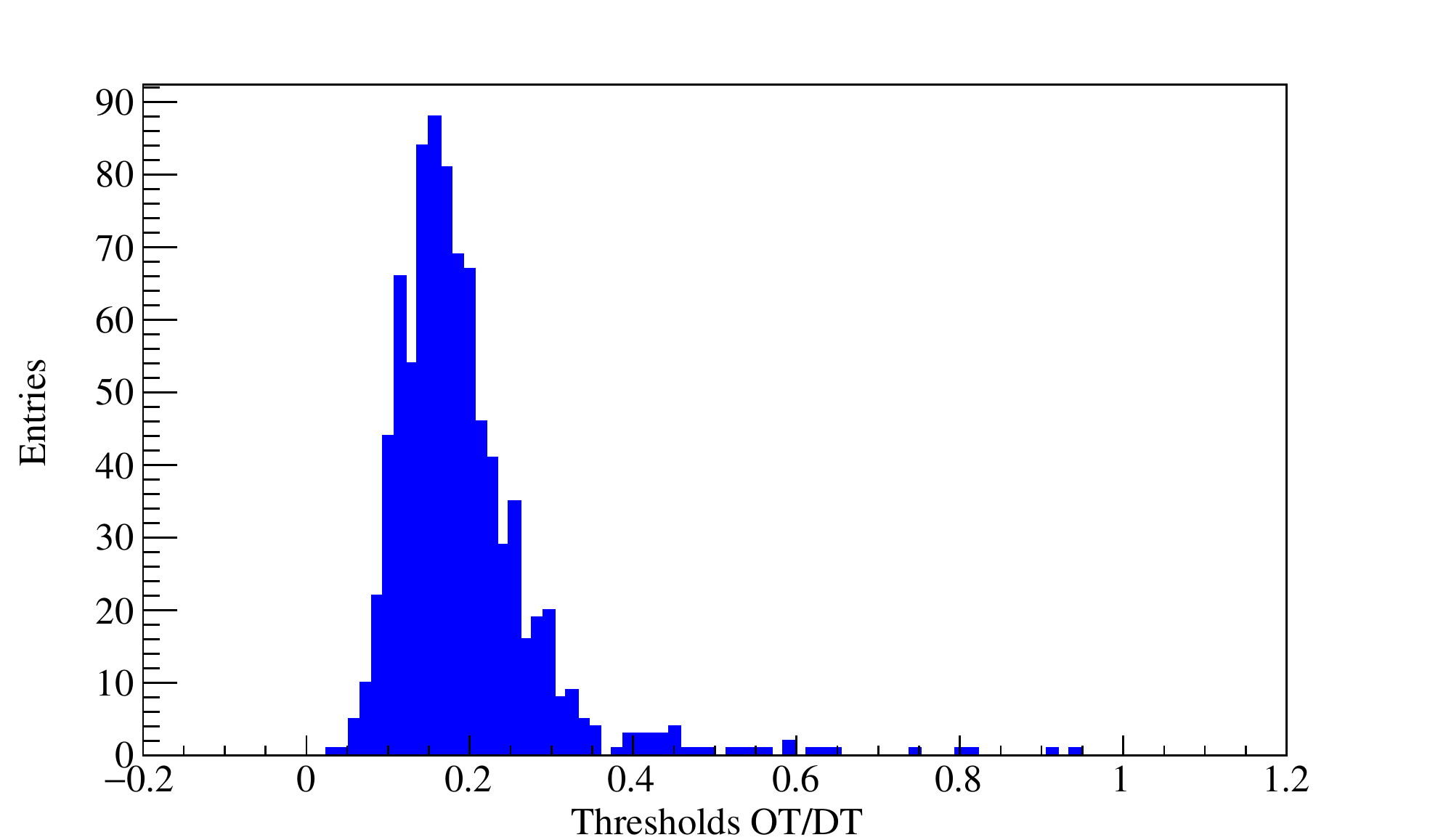}
\caption{
  Distribution of the ratio of OT and DT energy thresholds at 90\% of the trigger efficiency, evaluated using a dedicated run with pulser signals in a wide range of amplitudes.
  The histogram contains 858 entries, corresponding to almost 86.8\% of the CUORE bolometers.
  The excluded bolometers are those for which the test measurement did not allow to evaluate both the OT and DT thresholds unambiguously.}
\label{fig:fig5}
\end{center}
\end{figure}

Another application of \Apollo is its use on continuous data stored on disk from a previous data-taking, allowing to re-trigger the data and build the events from the new triggers. 
We refer to this application of the DAQ software as \Apollo re-triggering in the following.
\Apollo re-triggering works similarly to its online counterpart, but with a few differences and simplifications. 
In the first place, the software components that read the data from the digitizer boards when running online, are replaced in re-triggering by software components that instead read the continuous data files from disk.
Any parameter that depends only on the software configuration, can be changed during re-triggering.
The re-triggering consists often in a different trigger configuration or in a different length of the waveform window that is associated to correspondent trigger.
If better thresholds are indeed evaluated from the off-line analysis of the data, the derivative trigger can be run with updated thresholds.
Pulser events depend on the experimental setup, and for instance cannot be changed during re-triggering: the flags identifying a pulser event can only be replicated in the new data-stream.
Because the re-triggering does not need to interface to any hardware, it can be run offline on a computing cluster.
Therefore its speed is limited only by the available computing resources and it is usually much faster than the real-time data-taking. 
For example, with commonly available computing resources (CPUs running at speeds between 2 and 3\,GHz), the typical time needed to re-trigger a 24-hours CUORE run is about 3 hours with both OT and DT running in parallel.
The reduction of the processing time is possible not only because the \DataReaders do not need to wait for the data from the digitizers to become available, but also because the processing can be distributed over more CPUs and the pre-analysis algorithms that are usually run online are not executed in this case.
\Apollo re-triggering was extensively used in searches for dark matter~\cite{Alessandria:2012ha, DMCUORE-0, Alessandria:2012mt}, where the low energy spectrum of bolometric detectors is investigated.
The data used in these studies, initially acquired using the derivative trigger, were re-triggered with the OT algorithm in order to access the few keV range of the bolometers energy spectrum.
For instance, in CUORE-0 the OT allowed reaching trigger thresholds in the range 4--12\,keV, in contrast with the 30--120\,keV~\cite{Alduino:2016vjd} range achieved by the derivative trigger.
Currently, derivative trigger thresholds in CUORE range between $\sim$20 and 100\,keV, and a preliminary test of the OT on data shows that we can improve these values by 60--90\% (see Fig.~\ref{fig:fig5}).
The latest application of the re-triggering use of \Apollo was for the first CUORE dataset, used in \cite{Alduino:2017ehq}, that was first acquired with 5\,s waveform windows and then it was re-triggered at 10\,s after it was demonstrated that longer event windows help in increasing the signal to noise ratio.

\Apollo has been used also in conjunction with non-bolometric detectors. 
During the last 3 months of Cuoricino~\cite{Andreotti:2010vj}, the first full size prototype of a CUORE tower operated at LNGS from early 2003 to June 2008, the TeO$_2$ bolometers were acquired in coincidence with an array of plastic scintillators~\cite{Andreotti:2009dk}, with the purpose of studying the cosmic ray muon flux and the consequent induced event rate in the region of interest for 0$\nu\beta\beta$. 
Thanks to its underground location the soft cosmic ray component never get to the detector and the flux of penetrating muons is reduced by approximately six orders of magnitude.
However, the residual cosmic muon flux of 1.1$\times 10^{-6}$ muons/(h$\cdot$m$^2$) with an average energy of 270\,GeV~\cite{Ambrosio:1995cx}, can indeed interact in the detector or cause particle showers (mostly neutrons), resulting in events that can mimic a 0$\nu\beta\beta$ decay.
The scintillators were installed in appropriate positions in order to maximize the detector coverage, thus the probability of catching a muon event in coincidence with one of the Cuoricino crystals.
The scintillators read out was carried out using photomultipliers tubes (PMTs). 
The signals from the PMTs were handled by standard NIM and VME modules and eventually the information about the event time and charge collected from each PMT was propagated to an early version of \Apollo so that it could be integrated with the bolometer events.
The operation of the Cuoricino detector in coincidence with plastic scintillators allowed to obtain a single-bolometer background events upper limit of 0.0021\,counts/(keV$\cdot$kg$\cdot$yr) (95\% C.L.) in the region of interest for 0$\nu\beta\beta$ of $^{130}$Te. 
This level of background is negligible for CUORE, but it could be a limitation for the future CUPID interest group~\cite{Wang:2015raa} whose target background level is much lower than that of CUORE.
Consequently having a muon veto becomes crucial and with it the ability of \Apollo of handling also data from scintillators.

  We conclude this section with a discussion on the factors that could limit the usage of \Apollo in applications with requirements more stringent than those met for CUORE.
  We consider a setup with a larger number of acquired channels N, higher digitizers sampling frequency S or trigger rate R.
  The limitations could come from the data throughput and from the available CPU resources.
  For the data throughput we consider the amount of data read from the digitzers, then streamed over the local network and finally saved to files.
  In CUORE the data throughput, proportional to the product N$\cdot$S, is limited by the amount of data coming out form the NI digitizers.
  According to the manufacturer specification, for each digitizer board the product N$\cdot$S cannot exceed 16\,kSamples/s\footnote{This only applies if the anti-aliasing filter of the digitizer boards is active, however this is the required configuration in all uses of \Apollo}.
  In case of a whole chassis hosting 13 digitizers, this translates into $\sim$210\,kSamples/s or $\sim$6.5\,Mbit/s.
  On the other end of the connection, where the \Builder handles the data from all the six \DataReader, the data throughput is of the order of 40\,Mbit/s.
  This amount of data can be handled by any reasonably modern storage system, leading to the conclusion that at present \Apollo is not limited by the data throughput.
  Instead, in the case of CUORE, the performances are limited by the available CPU resources. 
  On the reader computers (Intel E3-1225 v3 4-core processor, running at 3.2\,GHz) the usage of CPU resources is dominated by the trigger algorithms.
  To make this statement more quantitative, we consider that in CUORE, where each reader computer has to handle about 200 channels, the OT online algorithm could not keep up with the data being acquired at 1\,kHz, so that we had to run it on data subsampled at 125\,Hz.
  On the event-builder computer (Intel E5-2620 v2 6-cores processor, running at 2.1\,GHz) the usage of CPU resources is dominated by the event building process (time-sort of the triggers, fetching of the associated waveforms from the circular buffers and pre-analysis algorithms run online).
  In CUORE this limits to a few hundreds Hz the maximum affordable trigger rate integrated over $\sim$1000 channels.
  These computational limits could be pushed back to some extent by using more performing hardware or by optimizing the software algorithms.
  However we note that both limits are related to triggering, an operation that needs not necessarily be performed online, provided that the continuous detector waveforms are saved by \Apollo.
  In particular, the scaling of \Apollo to a system more demanding than CUORE, would clearly benefit from moving the triggering and event building tasks offline.
  In this case R could be arbitrarily increased, while the limit on the product N$\cdot$S, coming from the NI digitizer boards, would still be present.
  If one wants to stick to the same digitizer boards used in CUORE, increasing S would imply a reduction in the channel density per board, and therefore per chassis and reader computer.
  Otherwise, a viable alternative would be to replace the digitizer boards, so as to remove at the source the limit on N$\cdot$S.

\section{Conclusion}
We described \Apollo, the data acquisition and control system of the CUORE and CUPID-0 experiments.
\Apollo was also used in many other cryogenic measurements, ranging from small R\&D setups with a few channels to the CUORE-0 experiment, thanks to the modularity, scalability and configurability of its hardware and software components.
One of the key features of \Apollo is its trigger-less data acquisition approach, which makes it especially suitable for experiments where saving the whole detector waveforms is preferable.
  Having access to the complete waveforms makes it possible to implement software triggers, with the appropriate complexity depending on the specific application.
In particular, throughout its applications \Apollo has demonstrated its ability to handle the data of a ton-scale experiment, and to deal with the readout of hybrid bolometers with dual heat and light signals.
Finally, it demonstrated its capability of integrating with other subsystems of the experimental apparatus, such as the PMT-based readout of a muon veto system based on plastic scintillators.
All of these characteristics make \Apollo well suited for a possible next-generation bolometric experiment aimed at the exploration of the parameter space corresponding to the inverted hierarchy region of the neutrino mass scale.

\acknowledgments

This work was sponsored by the Istituto Nazionale di Fisica Nucleare (INFN). 
In addition we would like to acknowledge the support from  the DOE Office of Science, Office of Nuclear Physics, Contract No. DE-FG02-08ER41551.
The authors thank the CUORE and CUPID-0 Collaborations, the directors and staff of the Laboratori Nazionali del Gran Sasso and the technical staff of Genova INFN division.
We would like to thank M.~Cariello for his great help in the design, assembly and test of the \Apollo hardware and his help during the DAQ installation of CUORE. 

\bibliographystyle{JHEP}
\bibliography{\jobname}

\end{document}